# A New Semi-Empirical Model for Cosmic Ray Muon Flux Estimation

Junghyun Bae[1]*, Stylianos Chatzidakis[1]
[1]School of Nuclear Engineering, Purdue University, West Lafayette IN 47906

**ABSTRACT**

Cosmic ray muons have emerged as a non-conventional high-energy radiation probe to monitor dense and large objects. Muons are the most abundant cosmic radiation on Earth, however, their flux at sea level is approximately 10,000 $min^{-1}m^{-2}$ much less than that of induced radiation. In addition, cosmic ray muon flux depends on not only various natural conditions (e.g., zenith angle, altitude, or solar activities) but also the geometric characteristic of detectors. Since the low muon flux typically results in long measurement times, an accurate estimation of measurable muon counts is important for muon applications. Here we propose a simple and versatile semi-empirical model to improve the accuracy in muon flux estimation at all zenith angles by incorporating the geometric parameters of detectors, and we name this the "effective solid angle model". To demonstrate the functionality, our model is compared with i) the cosine-squared, ii) PARMA model, and iii) Monte-Carlo simulations, and iv) measurements. Our results show that the muon count rate estimation capability is significantly improved resulting in increasing a mean C/E from 0.7 to 0.95. By selecting an appropriate intensity correlation, the model can be easily extended to estimate muon flux at various altitude and underground level.

- **Keywords**: Cosmic ray muons, Muon flux, Solid angle, Radiation detection, Instrumentation

*Corresponding author
Email address: bae43@purdue.edu

## 1. INTRODUCTION

Cosmic ray muons account for the major portion of cosmic radiation reaching Earth [1]. Due to their high-energy and penetrating nature, cosmic ray muons have been proposed as a non-conventional radiation probe for imaging and monitoring high-Z, large, and dense objects, typically a challenge with conventional radiography techniques. Specifically, muon applications have been studied in many fields, i.e., spent nuclear fuel cask and reactor imaging [2]–[8], homeland security [9]–[16], geotomography [17]–[20]. Despite efforts to further maximize the utilizability of cosmic ray muons [21]–[24], their applicability is often limited by the naturally low flux at sea level (about $10^4$ muons per minute in $m^2$) and its large variance with zenith angle. Cosmic muons are mainly produced by pion decays at an altitude of approximately 15 km. The measurable cosmic ray muon flux highly depends on the muon's traveling length because it decays to electron/positron with a mean lifetime of 2.2 $\mu$sec and it is attenuated in the Earth's atmosphere. One widely used empirical model for the relation between cosmic particle flux and zenith angle is the cosine-power model. For cosmic ray muons at sea level, the power of cosine is approximated to 2, or $I(\varphi) = I_0 \cos^2\varphi$ [25]. However, the cosine-squared model is limited due to its assumption for detector geometry, a point-detector. For example, according to the cosine-squared approximation, no muon will be detected when $\varphi = 90°$ regardless of detector size or configuration. However, cosmic muons can be measured at any zenith angle and there exist radiographic techniques using horizontal muon detectors [26], [27]. Although accurate analytical models for the terrestrial cosmic ray flux estimation are developed [28]–[30], detector geometry and configuration are not taken into account in these models. To address this gap, we aim to develop a new model that can easily estimate the expected cosmic muon count rates in all zenith angles at sea level for use in engineering applications.

In most muon applications including muon radiography and monitoring, at least two-fold coincidence muon detectors are installed and the target objects are placed in between two detectors. According to the cosine-squared model, the expected muon counts only depend on the active surface area of the detectors. However, they also depend on the distance between two detectors because their effective solid angle becomes larger as the distance decreases. To consider the geometry and configuration of detectors in the cosmic muon flux estimation, we developed a new approach named the effective solid angle. To demonstrate the performance of the effective solid angle model, we design and perform cosmic ray muon experiments and Monte-Carlo simulations. The experimental and simulation results are compared with both empirical and analytical estimations: cosine-squared, PARMA model [29], and effective solid angle models.

Another way to improve the cosine-squared model focuses on finding a better estimation for the power of cosine (instead of 2) [31], [32]. However, none of them provides a solution to the large uncertainties in high zenith angles using either cosine-power or -squared model. For muon applications, data acquisition and statistics are essential processes because they determine the quality of outcomes. Hence, the effective solid angle model can play an important role to estimate the necessary measurement time or volume of muon data to achieve the expected outcomes with small uncertainty in all zenith angles. In the end, we present the effective solid angle equations that can be universally applied in all detector configurations and zenith angles. In addition, a semi-empirical equation is included so that the computed effective solid angle is converted to muon count rates.

## 2. SEMI-EMPIRICAL EFFECTIVE SOLID ANGLE MODEL

Although the centerline of a detection system is aligned with the vertical direction, $\varphi = 0°$, it will not detect only muons that come at 0° zenith angle. Due to the finite size of the detector systems, some muons with $\varphi_\mu \neq 0$ will be detected. Similarly, even though the centerline of a detection system is aligned with the horizontal direction, $\varphi = 90°$, some muons are still detected because not all muons come from the horizontal direction. Therefore, both the detector geometry and distance between two detectors must be considered when estimating the muon flux.

The effective muon solid angle depends on both size and distance of two detectors and also it continuously varies on the detector surface as shown in Figure 1 (left). Because the muon flux variance along the azimuthal angle is insignificant [33], the solid angle only depends on the distance from the detector centerline on the surface. An example of the approximated solid angle of a point $P$ (red) which is away from the centerline with a radius, $r$, and the detector distance, $D$, is shown in Figure 1 (right). To compute the effective muon solid angle, we begin from finding the projected plane angle, $\vartheta$, which is the two-dimensional angle at point $P$. The projected plane angle is a function of $r$ whereas it is independent of the height of detector under the assumption that all muons are detected when they traverse detectors regardless of their deposited energy. The projected plane angle in radian is given by

$$\theta(r) = \operatorname{atan}\left(\frac{r_d + r}{D}\right) + \operatorname{atan}\left(\frac{r_d - r}{D}\right) \tag{1}$$

where $r_d$ is the radius of the detector surface. The area-averaged projected plane angle, $\vartheta_{avg}$, and half-projected plane angle, $\gamma$, over the detector surface becomes

$$\theta_{avg} = \frac{1}{\pi r_d^2} \int_{A_d} r\theta(r) dA_d \tag{2}$$

$$\gamma = \frac{\theta_{avg}}{2} = \frac{1}{2L}\left[\ln\frac{4L^2 + 1}{(L^2 + 1)^2} + \frac{1}{L}\operatorname{atan}(2L) + 2\left(L - \frac{1}{L}\right)\operatorname{atan} L\right] \tag{3}$$

$$L \equiv r_d/D \tag{4}$$

where $A_d$ is the detector surface area and $L$ is the ratio of $r_d$ to $D$. Because $\gamma$ depends on not only $r_d$ but also the $D$, a new parameter, $L$, is defined. The variance of $\gamma$ as a function of $D$ for various detector radii, 2.54, 5.08, and 7.62 cm using Eq (3) to (4) is shown in Figure 2. When two detectors are attached, $D = 0$ cm, the half-projected plane angle is 90° regardless of detector radius. On the contrary, it approaches 0° as $D$ increases.

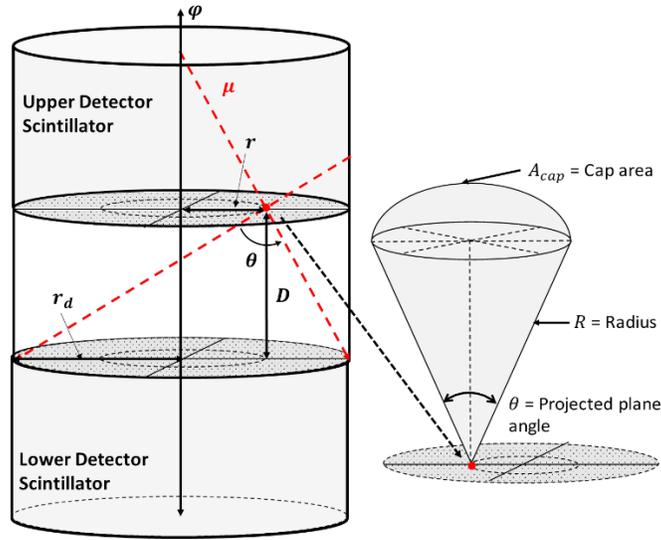

Figure 1 Example of cosmic ray muon trajectory (left) and the effective solid angle at the point on the detector surface (right).

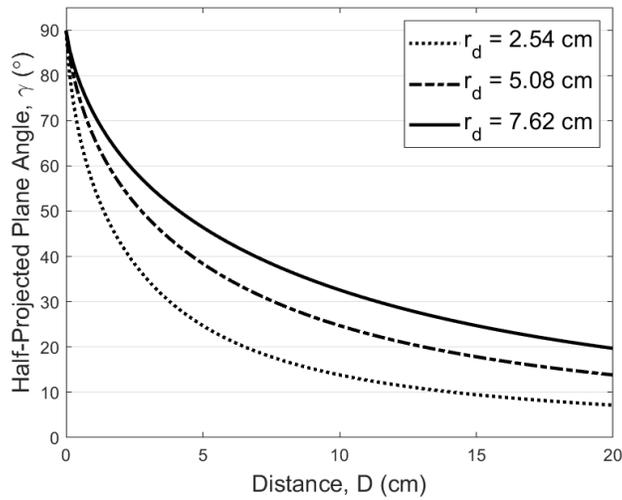

Figure 2 Half-projected plane angle as a function of the detector distance for $r_d$ = 2.54, 5.08, and 7.62 cm.

The expected muon flux varies within the solid angle which is confined to the detector size and distance. Since the azimuthal angular dependency to the muon flux is insignificant, when the half-plane projected angle is $\gamma$, the solid angle is

$$\Omega = 2\pi \int_0^\gamma \sin\phi \, d\phi \qquad (5)$$

The range of the pointing angle of a centerline, $\varphi$, extends from 0° to 90°. As the range of the pointing angle is uniformly divided by *N* number of angles

$$\varphi_i = \frac{\pi i}{2N}\bigg|_{i=1,2,\ldots,N} \tag{6}$$

By integrating the zenith angular dependent cosmic muon flux model, i.e., $I(\varphi)/I_0$ with Eq (5) and Eq (6), the effective solid angle over the entire azimuthal angle is given by

$$\Omega'_{\text{eff}}(i) = \frac{2\pi}{I_0} \int_{\varphi_i - \gamma}^{\varphi_i + \gamma} I(\phi) \sin \phi \, d\phi \tag{7}$$

Because the muon flux is assumed to be invariant over the azimuthal angle, a scaling factor, *F*, can be expressed by

$$F(i) = \frac{A_{\text{cap}}}{A_{2\pi}(i)} = \frac{1 - \cos \gamma}{\cos(\varphi_i - \gamma) - \cos(\varphi_i + \gamma)} \tag{8}$$

where $A_{cap}$ is the cap area shown in Figure 1 and $A_{2\pi}$ is a circular area between $\varphi \pm \gamma$. The complete effective solid angle equation for $i^{th}$ zenith angle becomes

$$\Omega_{\text{eff}}(i) = \Omega'_{\text{eff}}(i) F(i) \tag{9}$$

Since the expected cosmic muon flux is linearly proportional to the effective solid angle, a total muon count rate (CR) is given by

$$CR = \rho \Omega_{\text{eff}} \tag{10}$$

where $\rho$ is a conversion constant (muon count rate per steradian). The effective solid angle model as described in Eq (9) is independent of detector geometry and altitude. It can be solved either analytical or numerically as long as an appropriate intensity, e.g., cosine-power or cosine-squared, is selected. As an example, using the cosine-squared model which is the simplest cosmic ray muon flux estimation model, the effective solid angle model (Eq (7) and Eq (9)) becomes

$$\Omega'_{\text{eff}}(i) = \frac{2\pi}{3}[\cos^3(\varphi_i - \gamma) - \cos^3(\varphi_i + \gamma)] \tag{11}$$

$$\Omega_{\text{eff}}(i) = \left(\frac{2\pi}{3}\right)(1 - \cos \gamma)[(\cos(\varphi_i - \gamma) + \cos(\varphi_i + \gamma))^2 - \cos(\varphi_i - \gamma)\cos(\varphi_i + \gamma)] \tag{12}$$

The effective solid angle model is publicly available at MATLAB File Exchange [34].

## 3. EXPERIMENT SETUP

To benchmark the effective solid angle model, we performed a series of cosmic ray muon detection experiments using the two-fold coincidence measurement that consists of two independent scintillation detection systems (details of experimental environments are summarized in Table 1). In each detection system, a cylindrical sodium iodide scintillation crystal, NaI (Tl), integrated with a photomultiplier tube (PMT) was used to detect cosmic muons. The main specifications of NaI (Tl) crystal and photomultiplier base with preamplifier are summarized in Table 2. The average energy of cosmic ray muons at sea level is 3-4 GeV and the estimated muon energy deposition in the NaI (Tl) crystal using the Bethe equation is approximately 7 MeV/cm [35]. Although a total energy deposition depends on the muon traveling length in the crystal, the average amplitudes of scintillation photon signals are greater than most other background radiation signals. To minimize dead time caused by a long-tailed preamp pulse, we used an amplifier to transform the preamp pulse to the Gaussian-shaped short pulse. The amplifier gain is set as the minimum (× 2.5) in order to efficiently discriminate background noise. The reshaped signals are transmitted to the single channel analyzer which has a discriminator level of 10 V (maximum). Because the expected amplitudes of muon signals are greater than 12 V, most background noise signals ($\ll$ 10 V) are discriminated. To improve the muon detection reliability, we also used a coincidence logic gate which only accepts signals that occur within $500 \times 10^{-9}$ sec [36]. Using the coincidence measurement technique, we significantly minimized the random noise frequency [37].

Table 1 Experimental environments for cosmic ray muon flux measurements

| Date | Coordination | Elevation |
|---|---|---|
| May 2020 | 40°26'N, 86°54'W | 187 m |

Table 2 Selected specifications of NaI (Tl) scintillators, preamps, and PMT [38], [39]

| NaI (Tl) crystal | | Photomultiplier base with preamplifier | |
|---|---|---|---|
| Model | BICRON 2M2/2 | Model | ORTEC276 |
| Density | 3.67 g/cm$^3$ | PMT stages | 10 |
| Yields | ~38,000 photons/MeV | Conversion gain | $10^5$ - $10^6$ |
| Housing | Aluminum Thickness = 0.508 mm | Output rise/decay time | $t_r < 10^{-7}$ sec $\tau_d \approx 50$ $\mu$sec |
| Dimension | Diameter = 50.8 mm Height = 50.8 mm | Dimension | Diameter = 56 mm Height = 102 mm |

## 4. MONTE-CARLO SIMULATION

In addition to cosmic muon measurements, we performed Monte-Carlo simulation based on the cosine-squared model. Because a trace of muon trajectory on the detector surface is a random event, the possible interaction points are arbitrary generated on each detector surface (Figure 1). The Cartesian coordinates for muon traces on the upper and lower detector surfaces are

$$(x, y, z)_u = (x_m, y_m, D) \quad m = 1, 2, \ldots, N \tag{13}$$

$$(x, y, z)_l = (x_n, y_n, 0) \quad n = 1, 2, \ldots, N \tag{14}$$

where $N$ is the number of muon traces on each detector surface. By connecting two points from the upper and lower surfaces, $m$ and $n$, the three-dimensional muon trajectories are reconstructed. The reconstructed angles, $\vartheta_{m \to n}$, and expected muon fluxes, $I_{m \to n}$, are

$$\theta_{m \to n} = \arccos\left(\frac{D}{\sqrt{\Delta x^2 + \Delta y^2 + D^2}}\right) \tag{15}$$

$$I_{m \to n} = I_0 \cos^2(\theta_{m \to n}) \tag{16}$$

where $\Delta x = x_m - x_n$ and $\Delta y = y_m - y_n$. The total reconstructed angles, $\boldsymbol{\Theta}$, and intensities, $\boldsymbol{I}$, are given by

$$\boldsymbol{\Theta} = \begin{bmatrix} \theta_{1 \to 1} & \theta_{1 \to 2} & \cdots & \theta_{1 \to N} \\ \theta_{2 \to 1} & \theta_{2 \to 2} & & \theta_{2 \to N} \\ & \vdots & & \vdots \\ \theta_{N \to 1} & \theta_{N \to 2} & \cdots & \theta_{N \to N} \end{bmatrix} \tag{17}$$

$$\boldsymbol{I} = \begin{bmatrix} I_{1 \to 1} & I_{1 \to 2} & \cdots & I_{1 \to N} \\ I_{2 \to 1} & I_{2 \to 2} & & I_{2 \to N} \\ & \vdots & & \vdots \\ I_{N \to 1} & I_{N \to 2} & \cdots & I_{N \to N} \end{bmatrix} \tag{18}$$

The expected muon fluxes at $i^{th}$ zenith angle, $I_i$, and its mean intensity can be expressed as

$$\boldsymbol{I}_i = I_0 \cos^2(\boldsymbol{\Theta} \pm \varphi_i) \tag{19}$$

$$\bar{I}_i = \frac{1}{N^2} \sum I_i \tag{20}$$

## 5. RESULTS

To demonstrate the effective solid angle model, we compared with three results: i) the cosine-squared model, ii) a Monte-Carlo simulation, and iii) measurements. For the experiments, a total 21 sets of cosmic muon measurements were performed for seven zenith angles (0°, 15°, 30°, …, 75°, and 90°) and three distances (8, 9.5, and 11 cm) for 24 hours to minimize the day-night flux variation at sea level [40]. The results of muon measurements, estimations by the effective solid angle model and cosine-squared model, PARMA model, and Monte Carlo simulation are summarized in Table 3. The normalized quantities are also included in Table 3 in order to compare each other and those results are shown in Figure 3. In addition, a C/E (the ratio of Calculated and Experimental data) to the measured muon counts for the cosine-squared, effective solid angle models, PARMA model using EXPACS [30], and a Monte-Carlo simulation as a function of zenith angles are also shown in Figure 3. To eliminate the energy dependence, the results generated by EXPACS are integrated over muon energy (100 MeV to 1 TeV). It is noteworthy that the C/E for the cosine-squared model drops rapidly, especially, in high zenith angle levels (> 60°) because it assumes no muon is measured when $\varphi$ = 90°. On the contrary, both the effective solid angle model, PARMA model, and Monte-Carlo simulation are in good agreement with measurements. Although the C/E for the effective solid angle model decreases for high zenith angles due to low muon counts, it provides relatively stable predictions for all zenith angles (> 0.8). The Monte Carlo simulation shows the most accurate and stable prediction (1.0 ± 0.1). The C/E values for both effective solid angle and PARMA models constantly remain 1.0 ± 0.15 whereas that of the cosine-squared model decreases from 1 to 0. In addition, the mean C/E values for Monte Carlo simulation, effective solid angle, and cosine-squared models when the detector distances are 8, 9.5, and 11 cm, are shown in Figure 4. The accurate and constant C/E values of the effective solid angle model at all distances demonstrate the potential that it can be applied to various detector configurations.

Table 3 Experiment data (muon counts/day) and estimations by the effective solid angle model, normalized Monte Carlo simulation results, PARMA model, and cosine-squared model for zenith angles from 0° to 90° when $D$ = 8 cm and $r_d$ = 2.54 cm [41].

| Zenith Angle ($\varphi$) | Experiment Data | | PARMA [30] | Effective Solid Angle Model | Monte Carlo Simulation | $\cos^2\varphi$ |
|---|---|---|---|---|---|---|
| | **Counts/day** | **Normalized** | $\Phi_\mu/\Phi_0$ | $\Omega_{eff}/\Omega_0$ | $I_i/I_0$ | |
| 0° | 1929 | 1.0000±0.0322 | 1 | 1 | 1.0000±0.1000 | 1.000 |
| 15° | 1724 | 0.8937±0.0296 | 0.8680 | 0.9351 | 0.9368±0.0968 | 0.933 |
| 30° | 1472 | 0.7631±0.0264 | 0.7836 | 0.7577 | 0.7638±0.0874 | 0.750 |
| 45° | 1192 | 0.6179±0.0228 | 0.5573 | 0.5154 | 0.5397±0.0735 | 0.500 |
| 60° | 664 | 0.3442±0.0155 | 0.3626 | 0.2731 | 0.3121±0.0559 | 0.250 |
| 75° | 318 | 0.1649±0.0100 | 0.1942 | 0.0957 | 0.1467±0.0383 | 0.067 |
| 90° | 156 | 0.0809±0.0067 | 0.0567 | 0.0615 | 0.0876±0.0296 | 0.000 |

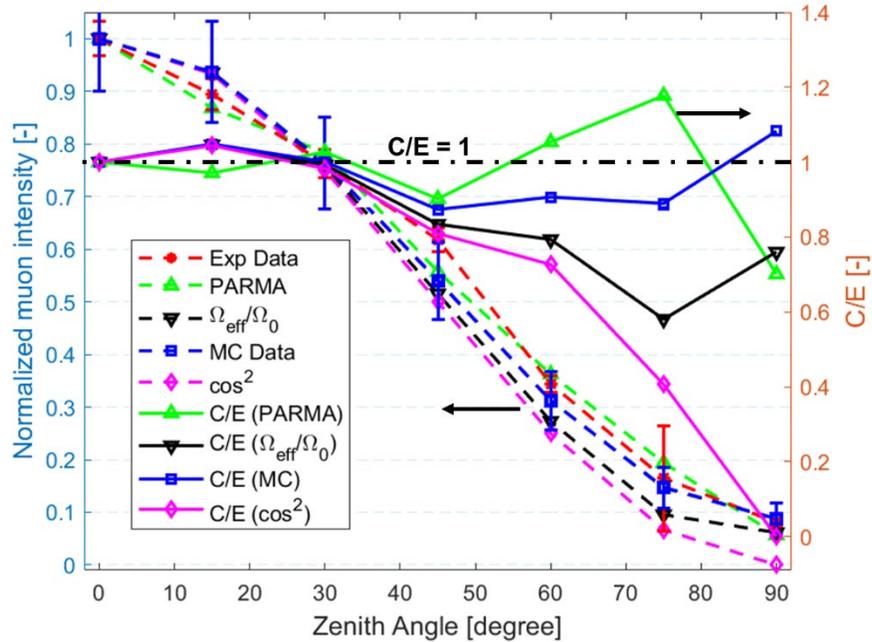

Figure 3 Comparison of cosmic muon measurements and four approaches: a) the effective solid angle model, b) cosine-squared model, c) PARMA model, and d) Monte-Carlo simulation (left y-axis) and their C/E (right y-axis) when $D$ = 8 cm and $r_d$ = 2.54 cm [41].

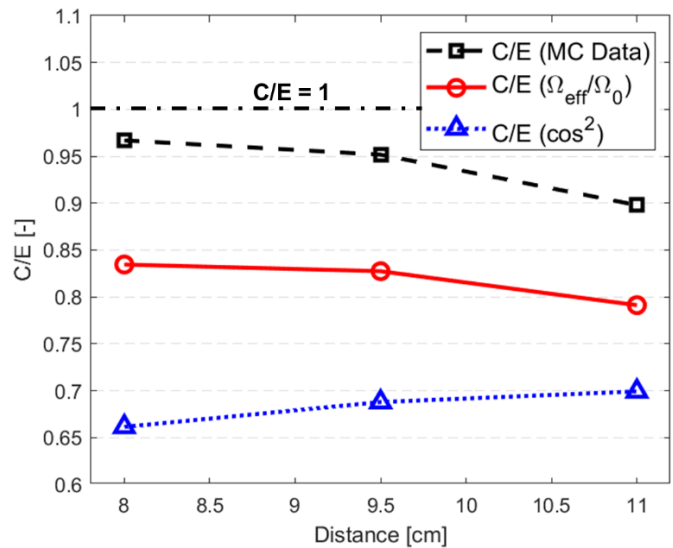

Figure 4 The mean C/E values as a function of detector distance, $D$, for three approaches: a) the effective solid angle model, b) cosine-squared model, and c) Monte-Carlo simulation.

The correlation between the daily cosmic muon counts and the effective solid angle is shown in Figure 5. The effective solid angle model accurately estimates the muon counts within a 1σ error level for $\varphi$ = 0° and 30°. On the other hand, it does not successfully estimate muon counts within a 1σ error level for $\varphi$ = 60° and 90°.

The effective solid angle model as a function of $L$ (Eq (3) and Eq (12)) and the measurement data when $r_d$ = 2.54 cm and $D$ = 8, 9.5, and 11 cm ($L$ = 0.3175, 0.2674, and 0.2309, respectively) are shown in Figure 6. In the range between $L$ = 0.15 and 0.4, both effective solid angle model and experiment data linearly increase as $L$ increases. Under conditions presented in Table 1, the semi-empirical correlation between the effective solid angle and muon count rate (CR) per day using the conversion constant, $\rho$, in Eq (10) is given by

$$CR \approx 7.52 \times 10^3 \left[\frac{\text{Counts/day}}{\text{sr}}\right] \times \Omega_{\text{eff}} \, [\text{sr}] \tag{21}$$

where $\Omega_{\text{eff}}$ can be found in Eq (9). Because the effective solid angle linearly increases when 0.2 < $L$ < 0.8, Eq (21) can be simplified by

$$CR \approx 7.52 \times 10^3 \times (CL + k) \tag{22}$$

The constants, $C$ and $k$, as a function of the zenith angle is shown in Figure 7 and summarized in Table 4 for selected zenith angles.

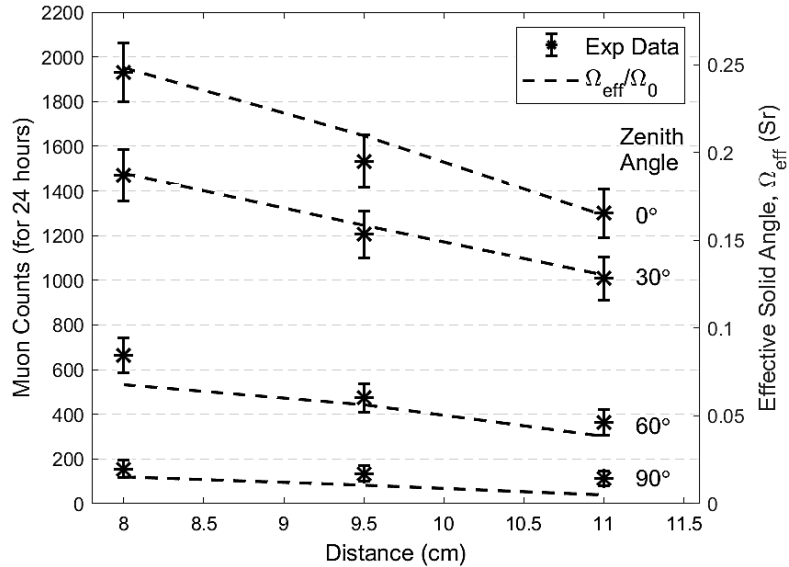

Figure 5 The effective solid angles as a function of the detector distance, $D$, for various zenith angles, 0°, 30°, 60°, and 90° and the cosmic muon measurement results (counts for 24 hours). The error bar represents 1σ.

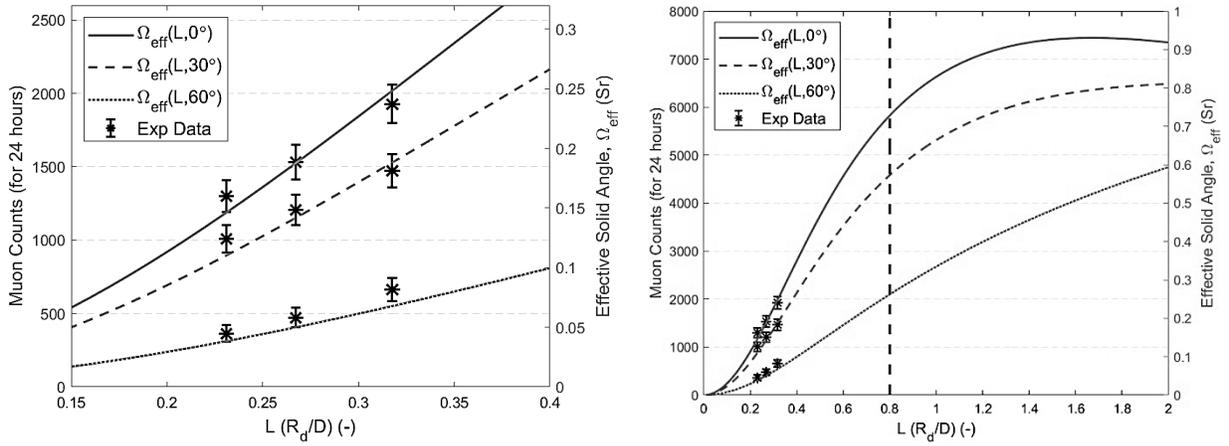

Figure 6 The effective solid angles as a function of $L$ for various zenith angles, 0°, 30°, 60° when 0.15 < L < 0.4 (left) and 0 < L < 2.0 (right). The effective solid angle equation can be simplified to the linear equation when 0.2 < L < 0.8.

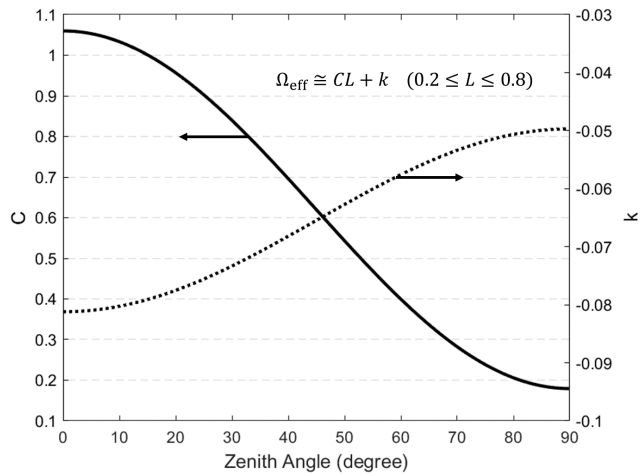

Figure 7 Constants, C and k, used in Eq (22) as a function of zenith angle when 0.2 < L < 0.8.

Table 4 Constants, C and k, used in Eq (22) for selected zenith angles when 0.2 < $L$ < 0.8

| Zenith Angles | C | k |
|---|---|---|
| 0° | 1.0599 | -0.0812 |
| 15° | 1.0009 | -0.0791 |
| 30° | 0.8398 | -0.0733 |
| 45° | 0.6197 | -0.0654 |
| 60° | 0.3995 | -0.0576 |
| 75° | 0.2384 | -0.0518 |
| 90° | 0.1794 | -0.0497 |

## 5. CONCLUSION

A new semi-empirical approach, the effective solid angle model, is developed to predict the cosmic ray muon flux at all zenith angles. By integrating the three-dimensional characteristic of muon detectors to the simple model (cosine-squared model), the effective solid angle model successfully estimates the cosmic ray muon count rates in all zenith angles. The model was compared with experimental measurements and we found that a mean C/E level is improved from 0.7 for the cosine-squared model to 0.95 for the effective solid angle model when D = 9.5 cm. It is worth noting that the model, despite its simplicity, is versatile enough to account for all detector geometries and configurations. The results can be improved by using more advanced model such as PARMA instead of cosine-squared model. The effective solid angle depends on two variables, detector surface radius and distance, and they are combined by introducing a new parameter, *L,* which is the ratio of the radius to distance. The effective solid angle model as a function of *L* shows linearity in which $0.2 < L < 0.8$ for all zenith angles. In addition, a semi-empirical conversion equation is presented so that the effective solid angle is converted to the cosmic ray muon count rate. The improved model to estimate the cosmic ray muon flux for all detection conditions is especially significant for high zenith angles (> 60°) because the cosine-squared model is limited in use for low zenith angles due to the large uncertainties. We anticipate our results to improve modeling quality in muon radiographic and monitoring applications by maximizing the utilizability of cosmic ray muons.

**ACKNOWLEDGMENTS**

This work was supported by the Purdue University Research Foundation.